\documentclass[useAMS]{mn2e}
\usepackage{rotating}
\usepackage{mncite}

\def\approxgt{\ifmmode \rlap{$>$}{}_{{}_{{}_{\textstyle\sim}}} \else%
$\rlap{$>$}{}_{{}_{{}_{\textstyle\sim}}}$\fi} 
\def\approxlt{\ifmmode \rlap{$<$}{}_{{}_{{}_{\textstyle\sim}}} \else%
$\rlap{$<$}{}_{{}_{{}_{\textstyle\sim}}}$\fi}

\LARGE \normalsize \title[]{Radio and X--ray observations during the
  outburst decay of the Black Hole Candidate XTE~J1908+094 }

\author[P.G. Jonker et al.]  {P.G. Jonker$^{1,2,3}$\thanks{email :
    pjonker@cfa.harvard.edu}, E. Gallo$^4$, V. Dhawan$^5$, M. Rupen$^5$, R.P. Fender$^4$, G. Dubus$^7$ \newauthor \\
  $^1$Institute of Astronomy, Madingley Road, CB3 0HA, Cambridge, UK\\
  $^2$Present address, Harvard--Smithsonian Center for Astrophysics, 60 Garden Street, MS83, MA 02138, Cambridge, U.S.A.\\
  $^3${\it Chandra} Fellow\\
  $^4$Astronomical Institute ``Anton Pannekoek'',
  University of Amsterdam, Kruislaan 403, 1098 SJ Amsterdam, The Netherlands\\
  $^5$National Radio Astronomy Observatory, Socorro, NM 87801, USA\\
  $^6$Laboratoire Leprince--Ringuet, Ecole Polytechnique 91128
  Palaiseau, France}

\begin{document}

\maketitle

\begin{abstract}
\noindent 
Obtaining simultaneous radio and X--ray data during the outburst decay
of soft X--ray transients is a potentially important tool to study the
disc -- jet connection. Here we report results of the analysis of
(nearly) simultaneous radio (VLA or WSRT) and {\it Chandra} X--ray
observations of XTE~J1908+094 during the last part of the decay of the
source after an outburst. The limit on the index of a radio -- X--ray
correlation we find is consistent with the value of $\sim0.7$ which
was found for other black hole candidates in the low/hard
state. Interestingly, the limit we find seems more consistent with a
value of 1.4 which was recently shown to be typical for radiatively
efficient accretion flow models. We further show that when the
correlation--index is the same for two sources one can use the
differences in normalisation in the radio -- X--ray flux correlation
to estimate the distance towards the sources if the distance of one of
them is accurately known (assuming black hole spin and mass and jet
Lorentz factor differences are unimportant or minimal).  Finally, we
observed a strong increase in the rate of decay of the X--ray flux.
Between March 23, 2003 and April 19, 2003 the X--ray flux decayed with
a factor $\sim$5 whereas between April 19, 2003 and May 13, 2003, the
X--ray flux decreased by a factor $\sim$750. The source (0.5--10 keV)
luminosity at the last {\it Chandra} observation was ${\rm L\approx
3\times10^{32} (\frac{d}{8.5 kpc})^2 erg\,s^{-1}}$.

\end{abstract}

\begin{keywords} stars: individual (XTE~J1908+094) --- stars: black holes
--- X-rays: stars 
\end{keywords}

\section{Introduction}
\label{intro}

Low--mass X--ray binaries (LMXBs) are binary systems in which a
$\approxlt 1\,M_{\odot}$ star transfers matter to a neutron star or a
black hole. These systems form one of our main windows on the physical
processes taking place around black holes and hence they can provide
us with information about the fundamental properties of spacetime. One
reason for this is that the great majority of Galactic black hole
candidates (BHCs) are found in transient LMXB systems.

Over the last few years it has become apparent that jets are an
integral and energetically important part of these BHC systems
(especially) when these systems are in the so called low/hard state
(\pcite{2001MNRAS.322...31F}; \pcite{2001MNRAS.327.1273S}).  Recently,
it was found that there exists a correlation between the radio and
X--ray flux in the low/hard state of several BHCs over 3--4 decades in
X--ray flux showing that there must be some form of disc--jet coupling
(\pcite{2003A&A...400.1007C}; \pcite{2003MNRAS.344...60G}).
\scite{2003MNRAS.343L..59H}, \scite{falcke2003}, and
\scite{merheinzdim2003} review the disc--jet connection in terms of
different accretion disc and jet models.  \scite{merheinzdim2003},
building on previous work of \scite{2003MNRAS.343L..59H}, showed that
inefficient accretion flow models can reproduce the observed radio --
X--ray correlation index for the initial parameter space they covered.
\scite{2003A&A...397..645M} showed that the jet--model explaining the
observed X--rays in terms of synchrotron emission from the jet of
\scite{2001A&A...372L..25M} can reproduce both the observed
correlation index as well as the normalisation.
\scite{2003MNRAS.343L..99F} used the observed radio -- X--ray
correlation for BHCs to argue that there is no need to advect energy
across a black hole event horizon in order to explain the observed
difference in quiescent luminosity between the neutron star and BHC
transient systems as was proposed by e.g.~\scite{2001ApJ...553L..47G}.
An important assumption in the work of \scite{2003MNRAS.343L..99F} is
that the observed radio -- X--ray correlation holds down to X--ray
luminosities as low as L$_X \sim10^{30-32}$ erg s$^{-1}$.

XTE~J1908+094 was discovered serendipitously during {\it RXTE}
observations of the soft gamma--ray repeater SGR~1900+14 by
\scite{2002IAUC.7856....1W}. The source flux is absorbed (${\rm
  N_H\sim2.3\times10^{22} cm^{-2}}$), the spectrum is well--fit with a
hard power--law with a photon index of 1.55.  Subsequent {\it
  BeppoSAX} observations (\pcite{2002IAUC.7873....1I};
\pcite{2002A&A...394..553I}) confirmed both the hard spectrum (the
source was detected up to 250 keV) and the high Galactic absorption. A
broadened iron emission line was present in spectra extracted from
both the RXTE and the BeppoSAX observations.  In 't Zand et al.
(2002) presented strong evidence for a low/hard -- high/soft state
change. The fact that the source displayed both a low/hard and a
high/soft state during the outburst is confirmed by the timing and
spectral analysis of the RXTE/PCA observations by
\scite{2002xrb..confE..11G} (see also \pcite{gogus2004}). A radio
counterpart was discovered by \scite{2002IAUC.7874....1R} whereas a
near--infrared counterpart was found by \scite{2002MNRAS.337L..23C}.
These authors also found that the optical upper limits
(\pcite{2002ATel...86....1W} and \pcite{2002IAUC.7877....4G}) are
fully consistent with the near--infrared colours of and the high
extinction towards the source.

In this paper we report the findings of our (nearly) simultaneous Very
Large Array (VLA)\footnote{The National Radio Astronomy Observatory is
  a facility of the National Science Foundation operated under
  cooperative agreement by Associated Universities, Inc.}  and
Westerbork Synthesis Radio Telescopes (WSRT)\footnote{The Westerbork
  Synthesis Radio Telescope is operated by the ASTRON (Netherlands
  Foundation for Research in Astronomy) with support from the
  Netherlands Foundation for Scientific Research NWO} radio and {\it
  Chandra} X--ray observations of XTE~J1908+094 during the last part
of the decay of the source after an outburst.

\section{Observations and analysis}
We have observed the BHC soft X--ray transient (SXT) XTE~J1908+094
using the ACIS detector in its Timed Exposure mode on--board the {\it
  Chandra} satellite (\pcite{1988SSRv...47...47W}) on three occasions.
Within a few days of the X--ray observations radio observations using
either the VLA or the WSRT were performed. A log of the observations
can be found in Table~\ref{log}.  All observing times have been
converted to UTC.

\subsection{The {\it Chandra} X--ray data}

The X--ray data were processed by the {\it Chandra} X--ray Center;
events with ASCA grades of 1, 5, 7, cosmic rays, hot pixels, and
events close to CCD node boundaries were rejected. We used the
standard {\it CIAO} software to reduce the data (version 2.3 and CALDB
version 2.21). A streak caused by the arrival of photons during the
CCD readout period (which lasts $\sim$41~ms in total) was present during
the first observation.

In all the observations we detect only one source. After applying the
{\it CIAO} web--based tool {\sc fix offsets} to correct for known
aspect offsets to each of the observations separately, we derive the
following coordinates for the source: R.A.=19h08m53.07s,
Decl.=+09$^\circ$23'05.0" (typical error 0.6", equinox 2000.0). The
radio coordinates of this transient (R.A.=19h08m53.077s
Decl.=+09$^\circ$23'04.9", with an error of 0.1", equinox 2000.0;
\pcite{2002IAUC.7874....1R}) are fully consistent with this.

\begin{figure*}
  \includegraphics[width=12cm,height=9cm]{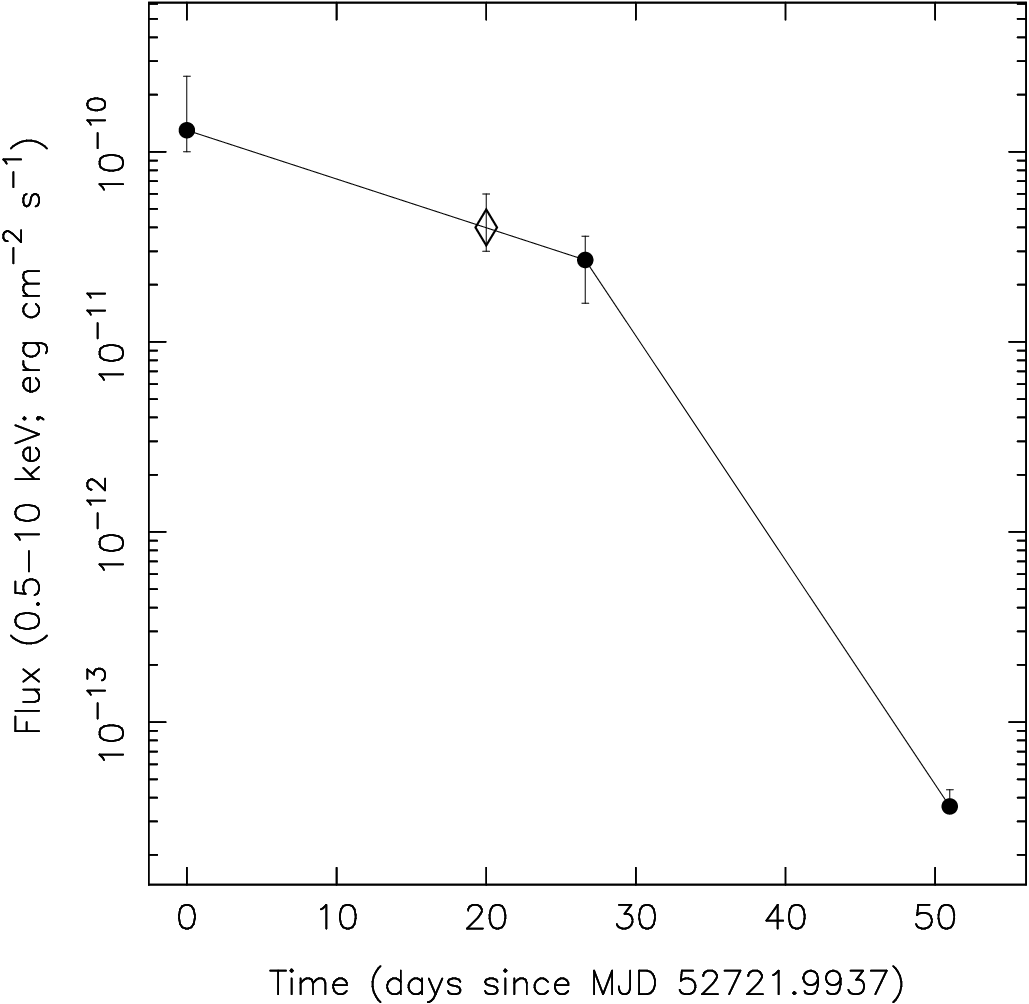}
\caption{The decay of the unabsorbed X--ray flux (0.5--10 keV) as a
  function of time.  Time zero is March 23, 23:51 UTC, 2003 (MJD
  52721.9937). The diamond is at the estimated X--ray flux of the
  source on April 12 assuming an exponential decay between March 23
  and April 19 (see text).}
\label{xdecay}
\end{figure*}

The source is detected at a count rate of 1.84$\pm$0.02 counts
s$^{-1}$ and 0.54$\pm$0.01 counts s$^{-1}$ for the March 23 and the
April 19 observation, respectively. These count rates have not been
corrected for effects of pile--up (see below). During the third
observation we barely detected the source; we detected 9.1$\pm$3.2
source counts (0.3--8 keV) spread over the entire length of the
observation (10.7 ksec, i.e.~a count rate of $\sim8.5\times10^{-4}$
counts s$^{-1}$). In Fig.~\ref{xdecay} we plot the three unabsorbed
X--ray fluxes (0.5--10 keV) as determined from spectral model fits to
the {\it Chandra} data (see below). The X--ray flux on April 12 was
estimated from an interpolation of the flux decay between March 23 and
April 19. We arbitrarily added a flux error of 50 per cent to this
point.  Between March 23 and April 19 the flux decayed by a factor of
$\sim$5, whereas between April 19 and May 13 the flux decayed by a
factor of $\sim$750. During the last observation the unabsorbed
0.5--10 keV source flux was $(3.6^{+0.8}_{-0.2})\times10^{-14}$ erg
cm$^{-2}$ s$^{-1}$. In this estimate we used a power--law index of
2.0$\pm0.5$, which is often found for quiescent BHCs
(\pcite{2002ApJ...570..277K}). This flux translates to a source
luminosity in the 0.5--10 keV band of ${\rm L\approx 3\times10^{32}
(\frac{d}{8.5 kpc})^2 erg\,s^{-1}}$.

The source spectra of the first two observations are extracted with 20
counts per bin. We only include energies above 0.3 and below 8 keV in
our spectral analysis since the ACIS Timed Exposure mode spectral
response is not well calibrated below 0.3 keV and above 8 keV. We fit
the spectra using the {\sc ISIS} package version 1.1.3
(\pcite{2000adass...9..591H}). Since the source is heavily absorbed we
hardly detected photons with energies below 1 keV. For this reason we
did not have to worry about the additional absorption due to
contamination by the optical blocking filters in our spectral fits
\footnote{see http://asc.harvard.edu/cal/Acis/Cal\_prods/qeDeg/}
(which primarily affects photons below 1 keV). However, we still used
the tool {\sc corrarf} to alter the auxiliary response file such that
this excess absorption is taken into account. Due to the high count
rate during the first observation the pile--up fraction is 25--30 per
cent for our frame time of 0.44~s (i.e.~a frame time of 0.4~s
since we had windowed the CCD to 1/8th of the nominal size plus a
deadtime of 41~ms for reading out the CCD). The pile--up fraction
during the second observation is approximately 10 per cent for the
same frame time.  In both cases we used the pile--up model of
\scite{2001ApJ...562..575D} to fit the spectra to model the effect of
pile--up on the spectrum. Furthermore, in order to test the results
from the pile--up model we extracted photons from the region covered
by the read--out streak which is unaffected by pile--up for the first
observation; two boxes, one $\sim$10x45 arcseconds$^2$ and the other
$\sim$10x15 arcseconds$^2$, were placed $\sim$5 arcseconds East and
West of the best--fit source position, respectively. The background
was determined from a region $\sim$10x60 arcseconds$^2$ placed
$\sim$16 arcseconds North of the best--fit source position.  We fitted
the spectrum obtained from the read--out streak using {\sc XSPEC}
(\pcite{ar1996}) version 11.3.0.

The effective exposure time for the readout streak data is much lower
than the actual exposure time since each pixel along the read--out
direction is illuminated by the source only for 40 $\mu$seconds (the
time necessary to transfer and read--out one row). Since the
streak pixels are illuminated at the approximately the same position
(with respect to on--axis) as the pixels during the frame integration
time we do not have to worry about differences in the effective area
and hence Auxiliary Response File for the streak and full frame
pixels. The effective streak exposure time for the streak boxes,
spanning $\sim130$ streak pixels as described above, is $\sim5.3\times
10^{-3}$~s per frame. For a total exposure time of 5.2~ksec a total of
approximately 5.2$\times 10^{3}$/0.44 frames are obtained. Hence, the
effective streak exposure time is $\sim$63~s.

\begin{figure*}
  \includegraphics[width=9cm,height=12cm,angle=-90]{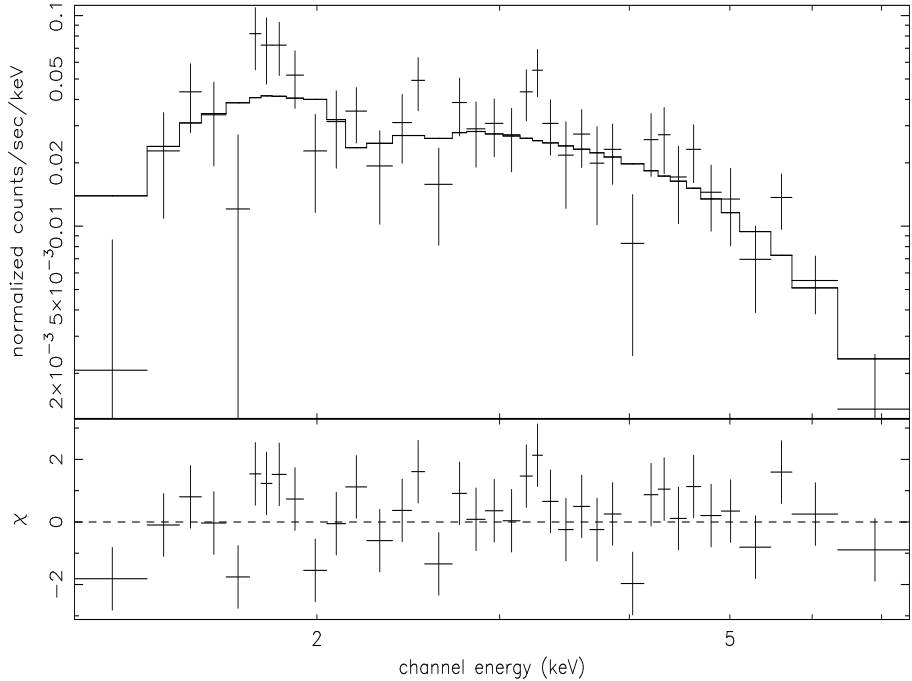}
\caption{{\it Top panel:} The 0.3--8 keV X--ray spectrum of
XTE~J1908+094 obtained with the {\it Chandra} observatory extracted
from the read--out streak (see text) on March 23, 2003. The solid line
is the best--fit absorbed power--law model. {\it Bottom panel:} The
residuals (data minus model) of the best--fit power--law model. }
\label{xspec}
\end{figure*}

Due to the pile--up the power--law index, normalisation, N$_H$, and
pile--up parameter $\alpha$ are degenerate. Indeed, a fit to the data
using an absorbed power law including the pile--up model gave an
unrealistically low value for N$_H$ of 1.6$\pm$0.5$\times10^{21}$
cm$^{-2}$ for the March 23 observation whereas N$_H$ was
4.0$\pm$0.3$\times10^{22}$ cm$^{-2}$ for the April 19 observation. The
N$_H$ is most likely close to 2.5$\times10^{22}$ cm$^{-2}$ found by
\scite{2002A&A...394..553I} using spectral modelling of BeppoSAX
outburst data of XTE~J1908+094. This is close to 1.8$\times10^{22}$
cm$^{-2}$ estimated using the model of \scite{1990ARA&A..28..215D}.
Therefore, during the fit we kept the value of the interstellar
absorption fixed at a value of 2.5$\times10^{22}$ cm$^{-2}$. An
absorbed blackbody model also represents the data well statistically
although the residuals show systematic trends and the pile--up
fraction needed to fit the second observation was unrealistically high
($\alpha\sim1$ whereas a value of 0.5 as found using the power--law
model is closer to the value expected based upon the pile--up
estimates). For that reason and since spectra of BHCs in the low/hard
state are usually well--fit with a power--law spectrum we only
included the parameters of the best--fit power--law model in
Table~\ref{specpars}. The fit result for the power--law index derived
using the read--out streak and pile--up model are consistent. To
estimate the uncertainty in the source flux we used the extrema of the
power--law indexes and normalisations encompassing the 90 per cent
confidence regions. The best--fit power--law spectrum of the first
observation extracted from the read--out streak is plotted in
Figure~\ref{xspec}.

\begin{table*}
\caption{Log of the radio and X--ray observations.}
\label{log}
\begin{center}

\begin{tabular}{lcccc}
\hline
Observation  & Calender Date   & Date MJD & Exposure time$^a$ & Band \\
 type  & UTC       & UTC & ksec.  &         \\
\hline
\hline
X--ray & March 23, 2003 at  23:51:00 & 52721.9937 & 5.2 & \\
Radio & March 25, 2003 at 13:00  & 52723.5416& 14.4 & 8.3 GHz\\
Radio & April 12, 2003 at 12:50  & 52741.5347& 14.4& 8.3 GHz\\
X--ray & April 19, 2003 at  14:57:16 & 52748.6231& 5.15 & \\
Radio & April 19, 2003 at 03:46 &52748.1569 & 43.2 & 5 GHz \\
Radio & May 13, 2003 at 13:17  &52772.5534 & 9 & 8.3 GHz\\
Radio & May 14, 2003 at 05:44  &52773.2388 & 25.2 & 8.3 GHz\\
X--ray & May 13, 2003 at  23:31:32 & 52772.9802& 10.7 & \\
\end{tabular}
\end{center}

{\footnotesize $^a$ For the radio observations the effective exposure
times are 60--70 per cent of the quoted times since calibrations are
performed intermittent the science exposures. For the X--ray
observations the effective exposure time is $\sim$90 per cent of the
quoted times since after a frame time of 0.4~s a deadtime for readout
of the frame of 41~ms is present.}\newline
\end{table*}

\begin{table*}
\caption{Best fit parameters of the spectra of XTE~J1908+094. All
quoted errors are at the 90 per cent confidence level. The local
absorption due to the {\it Chandra} optical blocking filters was
accounted for. The first line gives the best--fit results using the
spectrum extracted from the read--out streak. The X-ray flux is
corrected for interstellar absorption. }
\label{specpars}
\begin{center}

\begin{tabular}{lccccccc}
\hline
Observation  & N$_H$  & PL$^a$ & PL normalisation& Pile--up parameter& Flux (0.5--10 keV) &  Reduced \\MJD (UTC)
 & ($\times10^{22}$ cm$^{-2}$) &  Index  & photons keV$^{-1}$ cm$^{-2}$ s$^{-1,b}$  & $\alpha$& ergs cm$^{-2}$ s$^{-1}$  & $\chi^2/$d.o.f. \\
\hline
\hline
52721.9937 & 2.5$^c$ & 1.9$\pm$0.3 & (63$\pm$20)$\times$10$^{-3,d}$ & ... & 2.6$\times
10^{-10,d}$ & 1.17/36 \\
52721.9937 & 2.5$^c$ & 2.1$\pm$0.1 & (33$\pm$5)$\times$10$^{-3}$ &
0.8$\pm$0.2 & 1.3$^{+1.2}_{-0.3} \times 10^{-10}$ & 2.46/320    \\
52748.6231 & 2.5$^c$ & 1.3$\pm$0.2 & (2.5$\pm$0.3)$\times$10$^{-3}$ & 
0.5$\pm$0.2 & $(2.7\pm1.0)\times 10^{-11}$ & 0.95/106 \\

\end{tabular}
\end{center}

{\footnotesize $^a$ PL = power law}\newline {\footnotesize $^b$ Power
law normalisation at 1 keV.}\newline {\footnotesize $^c$ Parameter
fixed at this value during the fit.}\newline {\footnotesize $^d$ The
flux in the read--out streak and the PL normalisation are corrected
for the effective readout--streak exposure time (see text).}\newline
\end{table*}

\subsection{The VLA $+$ WSRT radio data}

We observed XTE~J1908+094 on four occasions at radio wavelengths.
Three VLA observations with the VLA in the D configuration at 8.3 GHz
and one WSRT observation at 5 GHz were obtained (see Table~\ref{log}).
The VLA radio data were reduced using the Astronomical Image
Processing System (AIPS), whereas the WSRT data was reduced using the
Multichannel Image Reconstruction, Image Analysis and Display package
(MIRIAD; \pcite{1995adass...4..433S}). A source was detected on two of
the four occasions at the position of the radio counterpart. The radio
flux was 0.355$\pm$0.018 mJy on March 25, 0.10$\pm$0.02 mJy on April
12, and 3 $\sigma$ upper limits of 0.1 mJy/beam on April 19 (WSRT
data, 5 GHz), and 57 $\mu$Jy/beam on May 14, 2003 on the source flux
at the position of the radio counterpart were obtained.  As mentioned
above the WSRT observations were made at 5 GHz whereas the VLA
observations were made at 8.3 GHz.  However, the radio spectrum of BHC
SXTs in the low--hard state is usually flat $\alpha\approxgt0$
(\pcite{2001MNRAS.322...31F}; S$_\nu\propto\nu^{\alpha}$). Hence,
assuming a flat radio spectrum the flux at 8.3 GHz will be the same as
that at 5 GHz.

\subsection{The X--ray -- radio correlation}

\begin{figure*}
  \includegraphics[width=12cm]{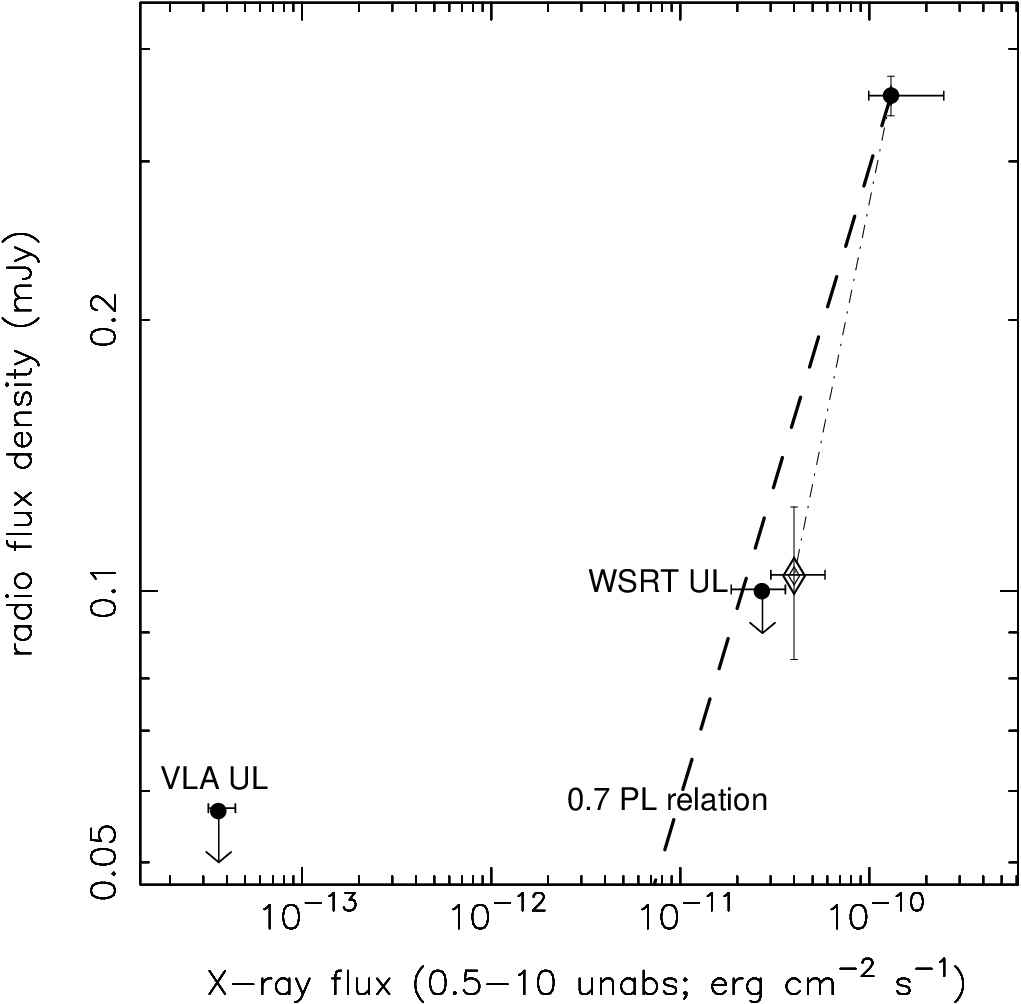}
\caption{Nearly simultaneous radio and X--ray observations of the BHC
   XTE~J1908+094. The power--law X--ray -- radio correlation with
   index 0.7 index as found before in other sources is indicated with
   the dashed line. The diamond is the point for which we estimated
   the X--ray flux (see text). The arrows indicate 3 $\sigma$ upper
   limits on the radio flux (VLA and WSRT). }
\label{xrrel}
\end{figure*}

We have one radio observation close in time (within two days) to the
X--ray observation on March 23, two radio observations which are
simultaneous with an X--ray observation (on April 19, and May 13,
2003), and one radio observation (on April 12, 2003) which was
obtained a week before an X--ray observation (see Table~\ref{log}). We
estimated the X--ray flux on April 12 by assuming an exponential decay
in X--ray flux with time between March 23 and April 19. However, if we
extrapolate the steep part of the decay (that between April 19 and May
13) backwards in time we would derive a much higher X--ray flux for
April 12. Furthermore, for several SXTs only the upper--envelope of
the decay follows the exponential decay profile, the flux on a
specific date can be lower than that predicted by an exponential decay
(\pcite{1997ApJ...491..312C}); hence the flux on April 12 could also
have been lower than our estimate. With these caviats in mind we
plotted the radio -- X--ray points in Figure~\ref{xrrel}. If we use
these points to put a limit on a radio -- X--ray correlation it seems
that the correlation is steeper than the previously observed
0.7$\pm$0.1 power--law index slope (denoted by the thick dashed line;
Gallo et al.~2003; \pcite{2003MNRAS.345L..19M}). If we assume that the
radio flux on April 19, 2003 was at the WSRT 1~$\sigma$ upper limit
and also take the March 23, 2003 radio/X--ray point, the power--law
index would be 1.5$^{+0.45}_{-0.3}$ (1~$\sigma$ errors, taking into
account the asymmetric errors in the X--ray flux).

Gallo, Fender, Pooley (2002, 2003) discuss reasons for the observed
spread in the normalisation constant of the radio -- X--ray
correlation. They show that the observed spread in the relation can be
described if the various sources have low Lorentz
factors. \scite{merheinzdim2003} showed that the normalisation depends
also on black hole mass and spin but for comparisons between X--ray
binaries this is likely to be less important. The main assumption in
the work cited above is that the physics responsible for the disc--jet
coupling is the same for each source. Below, we show that under this
assumption the observed spread in the correlation normalisation can
also be ascribed to the different distances towards the sources.
\begin{equation}
{\rm F_{Radio} \propto L_{Radio} \times d^{-2}}
\end{equation}
\begin{equation}
{\rm F_{Radio} \propto L_X^{0.7} \times d^{-2}}
\end{equation}
where we have used the ${\rm L_{Radio}\propto L_X^{0.7}}$ found by
\scite{2003A&A...400.1007C} and \scite{2003MNRAS.344...60G}. This
leads to
\begin{equation}
{\rm F_{Radio} \propto F_X^{0.7} \times d^{-0.6}}
\end{equation}
for the relation between the fluxes. Alternatively, if the radio--
X--ray correlation index is 1.4 instead of 0.7 then
\begin{equation}
{\rm F_{Radio} \propto F_X^{1.4} \times d^{+0.8}}.
\end{equation}

From these considerations it follows that a nearby source will be more
radio loud than a source at the same X--ray flux that is further away
if the relation between the radio and X--ray flux has an index smaller
than 1 whereas the reverse holds for systems with an index larger than
1. If the normalisation constant and distance are well known for one
source, we can in principle estimate the distance towards another
source which has the same index in the radio -- X--ray correlation
from the difference in the normalisations.

Assume we have measured the radio -- X--ray flux correlation for two
sources (both sources must have the same index). If the measured
normalisations of both relations are K$_1$ and K$_2$, the distance
d$_1=(\frac{K_1}{K_2})^{2-2b}\times d_2$, where b is the index of the
radio -- X--ray correlation.  Obviously, if there are differences in
the jet formation for the different sources the dependence of the
normalisation on source distance would be diluted.

\section{Discussion}

We have obtained (nearly) simultaneous VLA and WSRT radio and {\it
Chandra} X--ray observations of the BHC XTE~J1908+094 during the decay
after an outburst. We find that: \newline {\it (i)} Limits on a
power--law correlation between radio and X--ray flux suggest that the
power--law index may be larger than that found before in other
BHCs.\newline {\it (ii)} The rate of decay increases from a factor of
$\sim$5 in $\sim$25 days to a factor of $\sim$750 in $\sim$25
days. \newline {\it (iii)} The source spectrum hardens during the
decay. \newline Below we will discuss these findings in more detail.

Previously, \scite{2003A&A...400.1007C} and
\scite{2003MNRAS.344...60G} found that there is a correlation between
the radio and X--ray flux over more than 4 orders of magnitude. The
index of the power--law relation they fitted was consistent with being
the same for several sources at a value of 0.7$\pm$0.1 (Gallo et
al.~2003). If we assume that the radio flux of the source on April 19,
2003 was at the 1~$\sigma$ level of the WSRT upper limit the index was
1.5$^{+0.45}_{-0.3}$ for a two--point power--law decay for
XTE~J1908+094. This is consistent at the $\sim3\sigma$ level with the
value of 0.7 found before. However, there are some caveats. This index
was determined over a limited range in X--ray flux and using very few
measurements only. Furthermore, we assumed that the radio spectral
index is flat with $\alpha=0$ (S$_\nu\propto\nu^{\alpha}$). Finally,
the radio flux on Qpril 19, 2003 may have been much lower than what we
have assumed; this would make the correlation index steeper.

Previously, \scite{2003MNRAS.342L..67M} found an index of $\sim$1.4
for the radio -- X--ray correlation over a small range in X--ray flux
in the neutron star system 4U~1728--34. Possibly, XTE~J1908+094 is a
neutron star as well. However, \scite{2002A&A...394..553I} and
\scite{2002xrb..confE..11G} argue strongly in favour of a BHC nature
for XTE~J1908+094 on the basis of the observed outburst X--ray
spectral and timing properties (see also Gogos et al.~2004,
submitted), but obviously a dynamical mass estimate showing the mass
of the compact object to be more than 3 M$_\odot$ would settle the
issue.

If we follow the line of reasoning laid--out in
\scite{2003MNRAS.343L..99F} but taking ${\rm L_{Radio}\propto
  L_X^{1.4}}$ instead of ${\rm L_{Radio}\propto L_X^{0.7}}$ we find
that ${\rm L_{Jet}\propto L_X}$. Hence, the ratio between jet and
accretion power (which is assumed to be tracked by the X--ray
luminosity) remains the same as the source flux decays. We note that
the fact that the index was found to be 1.4 for a neutron star system
does not affect the conclusion of \scite{2003MNRAS.343L..99F} that the
difference in quiescent X--ray luminosity between BHCs and neutron
star soft X--ray transients can be explained without the need of
advection of energy across the event horizon as long as the index for
the BHCs is 0.7.

Recently, \scite{2003MNRAS.343L..59H} showed that the index of 0.7
follows naturally for several jet--models if one assumes that those
jet models are scale invariant (i.e.~the Schwarzschild radius, r$_s$,
is the only relevant length scale for jet formation).  Building on the
work of \scite{2003MNRAS.343L..59H}, \scite{merheinzdim2003} also
showed that other indexes for the radio -- X--ray correlations could
be found, i.e.~for both a gas and a radiation pressure dominated disc
the index would be $\sim$1.4, whereas it would be close to 0.7 for
radiatively inefficient accretion flows. \scite{2003A&A...397..645M}
show that the model explaining part of the X--ray emission as jet
synchrotron emission (\pcite{2001A&A...372L..25M}) can explain the 0.7
index as well as the normalisation of the radio -- X--ray correlation.

Hence, it seems that in XTE~J1908+094, during the part of the decay
that we covered with our radio and X--ray observations, a standard
geometrically thin optically thick disc plus a corona could have been
present. Why accretion would proceed via a geometrically thin disc in
XTE~J1908+094 whereas in other sources it is thought that the standard
disc is not present in the low/hard state is unclear. Perhaps it has
something to do with the luminosity levels at which the various
sources are observed so far, but since the distance to XTE~J1908+094
is ill--constrained (\pcite{2002A&A...394..553I} argue that the
distance must be larger than 3 kpc) the source luminosity is not well
known.

From Fig.~\ref{xdecay} it is clear that the rate of decay increased
enormously after April 12, 2003. Such a steep decrease has been
observed before for several BHCs and neutron star soft X--ray
transients (e.g.~\pcite{1997ApJ...491..312C}). In neutron star systems
this has been interpreted as evidence for the onset of the propeller
effect (\pcite{1998ApJ...499L..65C}), however, since such a drop in
luminosity seems to be common for BHCs as well, this interpretation
may need to be revised (\pcite{2003MNRAS.341..823J}).

Finally, we find that the X--ray spectrum hardens between the first
and second X--ray observation. Spectral hardening is often observed
during/just after a transition to the low/hard state
(cf.~\pcite{2001ApJ...563..229T}). However, the findings of Gogos et
al.~(2004) show that XTE~J1908+094 was already in the low/hard state
several months before the first {\it Chandra} observation was made.
Perhaps the source changed from the low/hard state back to a soft
state in between the {\it RXTE} and the {\it Chandra} observations.
Such a change would be consistent with the fact that the limit on the
radio -- X--ray correlation index is close to 1.4. We conclude that
more observations at these low flux levels (and likely low
luminosities) are necessary to determine the behaviour of these
sources when they return back to quiescence.

\section*{Acknowledgments} 
\noindent 
We would like to thank the {\it Chandra} Director, Harvey Tananbaum,
for approving these DDT observations, Yousaf Butt for help with
selecting the best observation modes, and the referee for useful
comments which helped improve the manuscript. PGJ was supported by EC
Marie Curie Fellowship HPMF--CT--2001--01308 and is supported by NASA
through Chandra Postdoctoral Fellowship grant number PF3--40027
awarded by the Chandra X--ray Center, which is operated by the
Smithsonian Astrophysical Observatory for NASA under contract
NAS8--39073. This research made use of results provided by the
ASM/RXTE teams at MIT and at the RXTE SOF and GOF at NASA's GSFC.

\end{document}